# Involving Users in the Design of a Serious Game for Security Questions Education

Nicholas Micallef and Nalin Asanka Gamagedara Arachchilage

Australian Centre for Cyber Security
School of Engineering and Information Technology
University of New South Wales
Canberra, Australia
e-mail: {n.micallef; nalin.asanka}@adfa.edu.au

## Abstract

When using security questions most users still trade-off security for the convenience of memorability. This happens because most users find strong answers to security questions difficult to remember. Previous research in security education was successful in motivating users to change their behaviour towards security issues, through the use of serious games (i.e. games designed for a primary purpose other than pure entertainment). Hence, in this paper we evaluate the design of a serious game, to investigate the features and functionalities that users would find desirable in a game that aims to educate them to provide strong and memorable answers to security questions. Our findings reveal that: (1) even for security education games, rewards seem to motivate users to have a better learning experience; (2) functionalities which contain a social element (e.g. getting help from other players) do not seem appropriate for serious games related to security questions, because users fear that their acquaintances could gain access to their security questions; (3) even users who do not usually play games would seem to prefer to play security education games on a mobile device.

## 1. Introduction and Background

Fall-back authentication mechanisms are used to recover forgotten passwords. Large organisations (i.e. Google, Facebook) have recently adopted fall-back authentication mechanisms such as text-based and email-based password recovery. However, these mechanisms do not manage to provide a better fall-back authentication experience because they are still prone to a number of security vulnerabilities and portability issues (Stavova et al. 2016). For instance, text-based password recovery has the limitation that users might not carry their device with them when on vacation. In this situation, it would be impossible for users to recover their forgotten password.

The main alternative to these fall-back authentication mechanisms are security questions. The main challenge of security questions is that strong answers to security questions (i.e. high entropy) are difficult for users to remember (Shay et al. 2012), but also hard for potential attackers to guess or obtain using social engineering techniques (e.g. monitoring of social networking profiles). Alternatively, weak answers to security questions (i.e. low entropy) are easier for potential attackers to breach (Bonneau et al. 2010; Denning et al. 2011), but then they are easier for users to remember (Zviran and Haga, 1990; Just and Aspinall, 2009, 2010). System-generated answers to security questions (as proposed by Micallef and Just (2011)) seem to be a promising design option (Micallef and Arachchilage, 2017a, 2017b), since they can limit the vulnerabilities to guessing and social engineering attacks (Shay et al. 2012). However, system-generated answers to security questions need to be better presented to users, to increase the strength of fall-back authentication. Hence, research still needs to investigate the best way to educate users to adhere to stronger answers to security questions.

A serious game is a game designed for a primary purpose other than pure entertainment (Djaouti et al. 2011). In the field of security education, serious games were used as a pedagogical tool, for teaching network security (Ariyapperuma and Minhas, 2005; Gondree et al. 2013), which has improved the overall learning experience. Also, previous research has used serious games in the security field to educate users about the susceptibility to phishing attacks, to teach users to be less prone to these security vulnerabilities (Arachchilage et al. 2013, 2014, 2016). Serious games were also used to motivate users to change their behaviour towards general concepts in computer security (Denning et al. 2013; Dasgupta et al. 2013) and to help users remember passwords (Tao and Adams, 2008; Malempati and Mogalla, 2011; McLennan et al. 2017). Most of this research was successful in motivating users to change their behaviour towards security issues through the use of serious games. Therefore, we argue that a serious game that nudges users to improve memorability towards security questions, through educational interventions, could be effective in enhancing users' behaviour to strengthen their answers to security questions.

A serious game for security questions has been recently proposed by Micallef and Arachchilage (2017a, 2017b), with the aim of enhancing the memorability of system-generated answers to security questions. However, the proposed game design was not empirically evaluated to investigate the features and functionalities that users would find desirable. Other previous HCI research has successfully involved users in the design of both be-spoke authentication mechanisms (Crawford et al. 2013, Micallef et al. 2015) and serious games for security education (Arachchilage et al. 2013, 2014, 2016). Hence, this research contributes to the field of security education for fall-back authentication by evaluating the design of the serious game proposed by Micallef and Arachchilage (2017a, 2017b), in a lab study. The aim of this research is to involve users in the design of this serious game, to investigate the features and functionalities that users find desirable to educate them to provide strong and memorable answers to security questions.

## 2. Game Features and Functionalities

To enhance the memorability of answers to security questions, Micallef and Arachchilage (2017a, 2017b) propose the use of the popular picture-based "*4 Pics 1 Word*" mobile game (Google play, 2014). This game provides players with challenges which requires them to find the word that relates the provided images

(e.g. for the 4 images in Figure 1a the relating word would be "Walk"). Micallef and Arachchilage (2017a, 2017b) use psychology research (Atkinson and Shiffrin, 1968) to justify their choice of adapting this game. They argue that the game's use of pictures and words could be used to create associations between cues and answers to security questions. This encoding through cue-association helps people remember and retrieve the stored information (i.e. answers to security questions) over a long period of time (Atkinson and Shiffrin, 1968). Micallef and Arachchilage (2017a, 2017b) do not evaluate the proposed game design. Hence, in this research we empirically evaluate how users perceive the design of this game to understand the features and functionalities that are required in a serious game that aims to educate users to provide strong and memorable answers to security questions. Next, we describe the main features and functionalities of the proposed serious game.

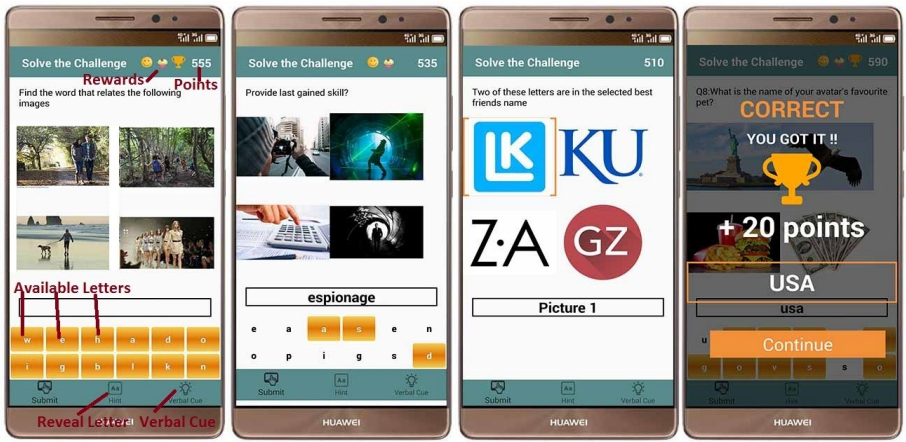

**Figure 1. Game challenges: a) standard, b) recall, c) recognition and d) badges**

**Challenges:** In most instances this game provides players with standard game challenges (as in the original "*4 Pics 1 Word*" mobile game). However, at certain intervals the game asks players to solve challenges related to the answers of the selected security questions (security questions challenges). The Generate-recognize theory (Anderson and Bower, 1972) states that recognition is easier and faster to perform than recall. Thus, for security questions challenges, the game was designed to use both recognition and recall challenges (see Figure 1b and 1c), because having only recognition challenges would have made the game more vulnerable to guessing attacks, since the answer space would have been very limited.

**Hints:** Anderson and Bower (1972) also found that it is difficult to remember information spontaneously without having any kind of memory cues. For this reason, the game was designed so that players are provided with 12 letters to assist them to solve a challenge (as in the original game). Besides showing the 12 letters, a feature was added to show verbal cues about each image. This feature is initiated by using the points that are gathered when solving other game challenges. Points can also be used to obtain hints to help solving more difficult challenges (i.e. deduction of 50 points for each hint, as in the original game). The more points players collect, the more hints they can obtain when they are struggling to solve other game challenges. To make the game less prone to guessing attacks, the game does not show the length of the word that needs to be guessed. This feature makes the game more difficult, but we argue that it makes the game less prone to guessing attacks.

**Points and Rewards:** Based on persuasive technology principles (Fogg, 2002), players can be conditioned to play a game if they are offered rewards to compensate their progress. Hence, in this game, players are awarded points when they solve challenges correctly (i.e. 10 for standard challenges, 15 for recognition security questions challenges and 20 for recall security questions challenges - these numbers were selected to condition players to put more effort in solving security question challenges). To understand whether rewards could also play an important role in enhancing intrinsic motivation (Deci et al. 1999) in serious games for security education, we reward players with badges (see Figure 1d) each time that they reach specific milestones related to security questions challenges.

## 3. Lab Study

A lab study was conducted to evaluate the design of a serious game that aims to educate users to provide strong and memorable answers to security questions. University ethics approval was obtained before conducting the study. The study consisted of two stages. The first stage was conducted separately because the game needed to be configured with the answers of the security questions that were selected by the participants. In this way, the evaluation of the serious game design was based on a realistic experience of using the game. In this stage participants, read the study information sheet, signed the consent form, selected security questions/answers and provided demographic information.

After configuring the game (this happened on the same day as Stage 1) participants were handed the answers to the security questions that they selected in Stage 1, to remind them of the answers that they provided. Afterwards, participants were handed a Samsung Galaxy S4 with the serious game described in the previous section and configured with the answers to the security questions that they selected in Stage 1. Participants played a test game together with the researcher, so that they could understand how to play the game. The test game never lasted more than 2 min. Then, they were told to play the game on their own using the provided mobile device.

The game started by selecting a random standard challenge (see Figure 1a) from a pool of 7 standard challenges. All players were provided the same 7 standard challenges, but in a random order. After completing a standard challenge, the player was given/deducted points (i.e. 10 points). Afterwards, the challenge was removed from the list of challenges. At this stage, the player was presented with a randomly selected recognition (see Figure 1c) security questions challenge (i.e. based on the security answers that were selected prior to playing the game). If the player picked the correct answer, together with points (i.e. 15 points), specific badges were rewarded (see Figure 1d), depending on the milestone that the player reached. The player was presented with alternate standard and recognition challenges until all 3 recognition security questions challenges were completed. After that, the player was prompted with alternate standard and recall challenges (see Figure 1b) until all 3 recall security questions challenges were completed. This is where the game ended. All players completed 13 challenges in total (7 standard challenges, 3 recognition security questions challenges and 3 recall security questions challenges).

After playing the game, participants also completed the System Usability Scale (SUS) questionnaire (Brooke, 1996), to measure their subjective satisfaction towards the mobile game interface design and its functionality. Afterwards, participants were asked to complete a post-study questionnaire, in which, they were asked about the game features, both current features and functionalities (explained in Section 2) and additional ones that could potentially be added to the game (e.g. leader board). Game performance was measured through log files.

### 3.1. Participants

20 participants (5 females, 15 males) were recruited through word of mouth and personal connections. The mean age was 29 (22-45), med=28. Most participants (16) were post-graduate students and the rest (4) were employed full-time. All participants (20/20) self-reported that they owned a smartphone for more than three years. Most of the recruited participants (12/20) were recruited because they did not play games regularly on their mobile devices and PCs, since we wanted to evaluate the serious game design even with users who do not play games regularly.

## 4. Results and Discussion

In this section we present and discuss the results extracted from the lab study described in Section 3. Our main findings reveal that: (1) even for security education games, features which provide rewards seem to motivate users to have a better learning experience; (2) functionalities which contain a social element (e.g. getting help from other players) do not seem appropriate for serious games related to security questions; and (3) even users who do not usually play games would seem to prefer to play security education games on a mobile device.

### 4.1. Current features and functionalities

Almost all participants (17/20) reported that they found the addition/deduction of points when solving challenges or requesting hints to be fair. Also, 18/20 participants liked the badges that were rewarded when solving security questions challenges. P8 said, "*It made me feel that I accomplished something*" and P13 said, "*Actually, it made me feel a little bit proud*". The only feature that was not well received was the verbal cues feature (participants were provided a verbal description of one of the four images) because only 2/20 participants used it. These results indicate that our participants appreciated the current game features and functionalities (see Section 2).

Previous research has found that rewards can promote feelings of competence which may increase intrinsic motivation (Deci et al. 1999). In the game-based learning for security education field, Arachchilage et al. (2013, 2014, 2016) defined a model which links the motivation of threat avoidance to rewards. However, this link has not been implemented and evaluated in practice. Hence, in our research we build upon this previous work by evaluating in practice the relationship between intrinsic motivation and rewards by awarding participants with badges when solving security questions challenges. Our findings revealed that these badges (see Figure 1d) were evaluated to be the most favourite feature of the game by almost all participants (18/20). Hence, one of our main findings is that we recommend the use of intrinsic rewards even for game-based learning for security education, since, we found that even in practice, intrinsic rewards seem to motivate users to have a better learning experience when using these technologies.

## 4.2. Additional features and functionalities

We also asked participants to provide us feedback about additional features and functionalities that could be added to the game. Most participants (17/20) would like to see progress about how many security question challenges they solved correctly. In most cases, they want to see this progress on a monthly basis (8/20), rather than weekly, or daily basis. Also, most participants (14/20) would like a new kind of hint feature that provides them with the number of letters that the word should contain. Most participants (13/20) would like a leader board in which they could compare their scores with other players. Most participants (11/20) would not like the feature of asking other players for help when they are struggling with solving standard challenges. Also, when asked whether they would like to select the images that appear in the game as cues to their answers, 12/20 reported that they would like the game to have this feature. These results indicate that participants' views were quite divided about most of these additional features and functionalities.

Denning et al.'s (2013) found that a social element can increase game engagement. However, our participants did not seem very keen to include social elements (i.e. help from other players to solve standard challenges and leader board functionality) in the game. They reported that they consider security questions to be private and did not want to risk that their acquaintances would gain access to their answers to security questions (although they were told that the proposed functionalities would not allow them to do so). Since our results are quite different from Denning et al.'s (2013) findings and social interactions have not been widely investigated in the field of security education, we suggest that social interactions should be investigated in more detail in games that try to educate people to enhance their security behaviour.

## 4.3. Game performance and experience

Our participants spent 7 minutes playing the game (fastest 3.5 min, slowest 10 min). They correctly solved 11/13 challenges (med=11, σ=±1.4), which were distributed as follows: (a) 5.7/7 standard challenges (med= 6, σ=±1.3); (b) 2.6/3 recognition (see Figure 1c) challenges (med=3, σ=±0.7); and (c) 2.7/3 recall (see Figure 1b) challenges (med=3, σ=±0.5). Participants used a mean of 4 hints (med=3, σ=±2.7) for standard challenges and only 2/20 participants used verbal cues. These results indicate that all participants had a consistent game playing experience.

At the end, we asked our participants to rate (on a 5 point Likert scale - from strongly disagree to strongly agree), whether the game was interesting and interactive. The participants found the game to be both interesting (mean=3.8, med=4, σ=±0.8) and interactive (mean=3.7, med=4, σ=±0.7). After playing the game, the experimenter evaluated participants' subjective satisfaction towards the mobile game interface design and its functionality using the SUS questionnaire. They evaluated the user satisfaction (usability) of the serious game interface and functionality (mean=65%, med=70%, σ=±15%) to be between "good" and "ok" (more closer to "good" - 71.4% than "ok" - 50.9%) based on Bangor et al's (2009) adjective scale rating of SUS scores. These results indicate that the first prototype of this game satisfied the participants because it was considered to be interesting and interactive.

In the game-based learning for security education field, Arachchilage et al. (2016) achieved a user satisfaction of 84% for their mobile game prototype. Thus, one could argue that if users are not satisfied with the game prototype design, the game would not motivate users to enhance their behaviour, which in our case it would not improve the memorability of strong answers to security questions. The detailed investigation of the user satisfaction results highlighted that the main weaknesses of the proposed game design interface and functionality were that most participants were still unsure on: (1) whether they would use the game very frequently; (2) whether the functions of the game were well integrated; and (3) whether they felt confident in using the game. To address these shortcomings and potentially improve the usability of the game, without affecting its' challenging aspect, the following improvements could be implemented: (1) adding more animations and sounds in order to fully immerse players into the real-world experience of the game; and (2) reduce the interaction steps required to solve a challenge, by adding other interactions, such as swipes and taps. Hence, these findings confirm Arachchilage et al. (2016) findings, which revealed that when designing tools to educate people, it is highly recommended to measure users' satisfaction because one cannot expect to have a better learning experience without a better interaction with the system.

**4.4. Potential adoption**

Participants were also asked whether they would download and play this game if it was freely available to download online (e.g. online market such as Android Play store). Most participants (13/20) reported that they would download and play this game. For instance, P1 said "*Yes because it made me feel relaxed and some tasks were challenging, which made it interesting*". Most of those participants that reported that they would not download the free app (4/7) motivated their choice by saying that they just do not play games on their mobile devices. For instance, P8 said "*I don't play games in my smartphone or tablet*". We also asked participants for which device should this game be developed (Micallef et al. 2016). Almost all participants (17/20) reported that they would like this game to be available on a mobile device (i.e. smartphone or tablet) rather than a computer (i.e. PC or laptop). For instance, P16 said "*It is easier to play games on my mobile phone*".

The potential adoption results presented in this section might be related to the user satisfaction results presented in Section 4.3. Hence, after that we add additional features and functionalities, we need to conduct further studies to investigate whether a more refined game prototype leads to a higher adoption and user satisfaction. From a device preference perspective, it is interesting to note that almost all users (17/20) reported that they would like this game to be implemented on a mobile device. Arachchilage et al. (2016) evaluated a serious game for phishing education on a mobile phone, however, they did not investigate whether users wanted to play the evaluated serious game on a mobile device. Hence, our research recommends that designers of serious games for security education may consider designing their games predominately for mobile devices (especially if their target audience is the 21-35 age group), because it seems that even users who normally do not play games (12/20 of our participants reported that they do not play games) reported that they were open to play a security education game on a mobile device.

## 5. Limitations and Future work

The main limitations of our evaluation are that we only evaluated the design of the serious game that aims to educate users to provide strong and memorable answers to security questions, with a small number of participants (20). Hence, future studies need to confirm whether the proposed serious game design helps improve long-term memorability of strong answers to security questions, with a larger number of participants (30+). Since another limitation of this work is that the SUS scores were not high enough, the next version of the game will be improved by: (1) adding more animations and sounds; and (2) reducing the interaction steps. We will also conduct a security evaluation of the proposed serious game to determine the security vulnerabilities that need to be addressed to achieve the required security level. Afterwards, we will conduct a longitudinal field study to determine how much training is required, for users to learn the answers to their security questions.

## 6. Conclusions

The aim of this research was to involve users in the design of a serious game, to investigate the features and functionalities that users find desirable to educate them to provide strong and memorable answers to security questions. Our findings reveal that our participants: (1) found most of the current features and functionalities (see Section 2) to be desirable; and (2) did not seem keen on having functionalities which contain a social element (e.g. getting help from other players), because they fear that their acquaintances could gain access to their security questions. Also, this research recommends: (1) the use of intrinsic rewards when designing games for security education, since, features which involve rewards seem to motivate users to have a better learning experience when using these technologies; and (2) that designers of serious games for security education may consider designing their games mainly for mobile devices (especially if their target audience is the 21-35 age group), because it seems that even users who usually do not play games would seem to prefer to play a security education game on a mobile device. In our future work, we will conduct a longitudinal field study to evaluate whether the proposed serious game improves the long-term memorability of strong answers to security questions.

# 7. References

Anderson, J. R. and Bower, G. H. (1972), "Recognition and retrieval processes in free recall." *Psychological review*, 79(2), 97.

Arachchilage, N. A. G. and Love, S. (2013), "A game design framework for avoiding phishing attacks." *Computers in Human Behavior*, 29(3), 706-714.

Arachchilage, N. A. G. and Love, S. (2014), "Security awareness of computer users: A phishing threat avoidance perspective." *Computers in Human Behavior*, 38, 304-312.

Arachchilage, N. A. G., Love, S. and Beznosov, K. (2016), "Phishing threat avoidance behaviour: An empirical investigation." *Computers in Human Behavior*, 60, 185-197.

Ariyapperuma, S. and Minhas, A. (2005), "Internet security games as a pedagogic tool for teaching network security." *Proceedings of FIE 2005* (pp. S2D-1). IEEE.

Atkinson, R. C. and Shiffrin, R. M. (1968), "Human memory: A proposed system and its control processes." *Psychology of learning and motivation*, 2, 89-195.

Bangor, A., Kortum, P. and Miller, J. (2009), "Determining what individual SUS scores mean: Adding an adjective rating scale." *Journal of usability studies*, 4(3), pp.114-123.

Bonneau, J., Just, M. and Matthews, G. (2010), "What's in a name? evaluating statistical attacks on personal knowledge questions." *Proceedings of International Conference on Financial Cryptography and Data Security* (pp. 98-113). Springer Berlin Heidelberg.

Bonneau, J., Bursztein, E., Caron, I., Jackson, R. and Williamson, M. (2015), "Secrets, lies, and account recovery: Lessons from the use of personal knowledge questions at google." *Proceedings of WWW 2015* (pp. 141-150). ACM.

Brooke, J. (1996), "SUS-A quick and dirty usability scale. Usability evaluation in industry", 189(194), pp.4-7.

Crawford, H., Renaud, K. and Storer, T. (2013), "A framework for continuous, transparent mobile device authentication." *Computers & Security*, 39, pp.127-136.

Dasgupta, D., Ferebee, D. M. and Michalewicz, Z. (2013), "Applying puzzle-based learning to cyber-security education." *Proceedings of InfoSecCD'13*, (p. 20). ACM.

Deci, E., Koestner, R. and Ryan, R. (1999), "A meta-analytic review of experiments examining the effects of extrinsic rewards on intrinsic motivation." *Psychological Bulletin* 125, 6 (1999), 627–668.

Denning, T., Bowers, K., Van Dijk, M. and Juels, A. (2011), "Exploring implicit memory for painless password recovery." *Proceedings of CHI 2011* (pp. 2615-2618). ACM.

Denning, T., Lerner, A., Shostack, A. and Kohno, T. (2013), "Control-Alt-Hack: the design and evaluation of a card game for computer security awareness and education." *Proceedings of the 2013 conference on Computer & communications security* (pp. 915-928). ACM.

Djaouti, D., Alvarez, J. and Jessel, J.P. (2011), "Classifying serious games: the G/P/S model." *Handbook of research on improving learning and motivation through educational games: Multidisciplinary approaches*, 2, pp.118-136.

Fogg, B. J. (2002), "Persuasive technology: using computers to change what we think and do." *Ubiquity*, 2002(December), 5.

Gondree, M., Peterson, Z.N. and Denning, T. (2013), "Security through play." *IEEE Security & Privacy*, 11(3), pp.64-67.

Google Play. (2014), "4 Pics 1 Word" https://play.google.com/store/apps/details?id=de.lotum.whatsinthefoto.us&hl=en Retrieved 10 July, 2017.

Just, M. and Aspinall, D. (2009), "Personal choice and challenge questions: a security and usability assessment." *Proceedings of SOUPS 2009* (p. 8). ACM.

Just, M. and Aspinall, D. (2010), "Challenging challenge questions: an experimental analysis of authentication technologies and user behaviour." *Policy & Internet*, 2(1), 99-115.

Malempati, S. and Mogalla, S. (2011), "An Ancient Indian Board Game as a Tool for Authentication." *International journal of Network Security and It's Applications, IJNSA*, 3(4), 154-163.

McLennan, C. T., Manning, P. and Tuft, S. E. (2017), "An evaluation of the Game Changer Password System: A new approach to password security." *International Journal of Human-Computer Studies*, 100, 1-17.

Micallef, N. and Arachchilage, N. A. G. (2017), "A Gamified Approach to Improve Users' Memorability of Fall-back Authentication." *Proceedings of SOUPS 2017*. USENIX Association.

Micallef, N. and Arachchilage, N. A. G. (2017), "Changing users' security behaviour towards security questions: A game based learning approach." *Proceedings of MilCIS 2017*. IEEE.

Micallef, N., Baillie, L. and Uzor, S. (2016). "Time to exercise!: an aide-memoire stroke app for post-stroke arm rehabilitation." *Proceedings of MobileHCI 2016* (pp. 112-123). ACM.

Micallef, N. and Just, M. (2011), "Using Avatars for Improved Authentication with Challenge Questions." *Proceedings of SECURWARE 2011*.

Micallef, N., Just, M., Baillie, L., Halvey, M. and Kayacik, H. G. (2015), "Why aren't users using protection? investigating the usability of smartphone locking." *Proceedings of Mobile HCI 2015* (pp. 284-294). ACM.

Shay, R., Kelley, P.G., Komanduri, S., Mazurek, M.L., Ur, B., Vidas, T., Bauer, L., Christin, N. and Cranor, L.F. (2012), "Correct horse battery staple: Exploring the usability of system-assigned passphrases." *Proceedings of SOUPS 2012* (p. 7). ACM.

Stavova, V., Matyas, V. and Just, M. (2016), "Codes v. People: A Comparative Usability Study of Two Password Recovery Mechanisms." *Proceedings of IFIP International Conference on Information Security Theory and Practice* (pp. 35-50). Springer International Publishing.

Tao, H. and Adams, C. (2008), "Pass-go: A proposal to improve the usability of graphical passwords." *IJ Network Security*, 7(2), 273-292.

Zviran, M. and Haga, W.J. (1990), "User authentication by cognitive passwords: an empirical assessment." *Proceedings of Information Technology, 1990.'Next Decade in Information Technology'*, (Cat. No. 90TH0326-9) (pp. 137-144). IEEE.